# Shared Autonomous Electric Vehicle Service Performance: Assessing the Impact of Charging Infrastructure and Battery Capacity

Reza Vosooghi[a,b,*], Jakob Puchinger[a,b], Joschka Bischoff[c], Marija Jankovic[b], Anthony Vouillon[d]

[a]*Institut de Recherche Technologique SystemX, Palaiseau 91120, France*
[b]*Laboratoire Génie Industriel, CentraleSupeléc, Université Paris-Saclay, Gif-sur-Yvette 91190, France*
[c]*Transport System Planning and Transport Telematics, Technische Universität Berlin, Germany*
[d]*Direction de la Recherche / Nouvelles Mobilité (DEA-IRM), Technocentre Renault, Guyancourt 78280, France*

ABSTRACT

Shared autonomous vehicles (SAVs) are the next major evolution in urban mobility. This technology has attracted much interest of car manufacturers aiming at playing a role as transportation network companies (TNCs) in order to gain benefits per kilometer and per ride. The majority of future SAVs will most probably be electric. It is therefore important to understand how limited vehicle range and the configuration of charging infrastructure will affect the performance of shared autonomous electric vehicle (SAEV) services. We aim to explore the impacts of charging station placement, charging type (rapid charging, battery swapping) as well as vehicle range onto service efficiency and customer experience in terms of service availability and response time. We perform an agent-based simulation of SAEVs across the Rouen Normandie metropolitan area in France. The simulation process features impact assessment by considering dynamic demand responsive to the network and traffic.

Research results suggest that the performance of SAEVs is strongly correlated to the charging infrastructure. Importantly, faster charging infrastructure and optimized placement of charging locations in order to minimize distances between demand hubs and charging stations result in a higher performance. Further analysis indicates the importance of dispersing charging stations across the service area and how this affects service effectiveness. The results also underline that SAEV battery capacity has to be carefully selected to avoid the overlaps between demand and charging peak times. Finally, the simulation results show that by providing battery swapping infrastructure the performance indicators of SAEV service are significantly improved.

## 1. Introduction

Shared autonomous vehicles (SAVs) are expected to be an integral part of future transportation systems playing an increasing role in the next five to 30 years (Greenblatt and Shaheen, 2015; Litman, 2018). Automakers who are ready to put the first fully autonomous vehicles (AVs) on the market in the near future, have announced their plan for ridesharing services with their self-driving cars in order to propose new beneficial solutions. Simultaneously, electric vehicle (EV) production continues its expansion ensuring the reduction of local pollutant emissions. Given important advances in battery technologies for EVs in recent years as well as the growing deployment of policy for achieving a shift towards electric and green transportation, these vehicles are forecasted to make up to as much as 30% of global auto production by 2030 (International Energy Agency., 2018). Considering these parallel evolutions, it is very likely that future SAVs, as a major part of current AV concepts, will mainly be powered by electricity. There are many reasons that make plausible future SAVs to be electric. First, the price of EV technology continues to fall and they become financially advantageous in comparison to vehicles with combustion engines (Berckmans et al., 2017; Nykvist and Nilsson, 2015). Second, EVs are best suited to reduce emissions in the sector and therefore help meet policy targets. Specifically in this case, it is suggested that autonomous EVs produce dramatically fewer emissions than gasoline AVs (Gawron et al., 2018) and consume fewer energy (Vahidi and Sciarretta, 2018). Third, EVs are technically and economically beneficial rather for longer daily travel distances experienced by shared fleets due to their relatively low maintenance needs (Logtenberg et al., 2018; Palmer et al., 2018; Weldon et al., 2018). However, despite the aforementioned advantages, configuration of a shared service based on EVs meets some operational concerns. Owing to the limited battery capacities and the lengthy charging process, a shared autonomous electric vehicle (SAEV) system may not achieve the same service usage than of a non-electric SAV system. Besides, providing charging stations can be very costly, specifically in the congested and high-density areas. Furthermore, charging outlets at each station are limited according to the available space and charging power, and can only be used for a small part of a fleet at a time. Thus, SAEVs' specification and charging station configuration must be carefully adjusted to meet the optimum service efficiency. It should be noted that the infrastructure needed for SAEVs may be substantially different from ordinary EVs (Weiss et al., 2017). Importantly, SAEV fleets could have a significant demand for rapid charging, potentially at high service demand areas and peak hours. Moreover, given the fastest charging rates provided by today's commercially available superchargers, it seems that a part of SAEV fleet will be unavailable for at least one hour per vehicle and per charge. This decrease of service availability may result in higher traveler wait times, and consequently lower demand and vehicle utilization. Thus, charging processes must be wisely scheduled to meet users' maximum demand.

---
[*] Corresponding author. Tel.: +33-6-6597-4453.
 *E-mail address:* reza.vosooghi@irt-systemx.fr

The SAEV vehicle specification (i.e. battery capacity or vehicle range) and the configuration of required infrastructure including the charging station placement, charging speeds and available spaces in charging stations could certainly affect service performance. These aspects have attracted less attention particularly when such services are simulated employing more sophisticated demand modeling and especially multimodal dynamic demand approaches. The purpose of this study is to provide a new insight into the design of SEAVs exploring how the service configuration could affect its effectiveness. With this aim, this research makes four major contributions. First, we propose new strategies of charging station placement and compare them. These strategies are based on two different optimization models. Second, we investigate for the first time the application of battery swapping stations (BSS) for SAEVs service. Third, we evaluate the performance of service according to the variation of vehicle/outlet ratio, which has not been until now the subject of investigations. Finally, a real world case study, based on the population and trip patterns of the Rouen Normandie metropolitan area in France is employed to demonstrate impacts of charging infrastructure and SAEV battery capacity on the service performance and its effectiveness. In order to carry out these investigations, an activity-based multi-agent simulation is used. The simulations incorporate dynamic traffic assignment in which SAEV mode choice is integrated in multimodal travel demand patterns according to user taste variations.

The remainder of this paper is structured as follows. In section 2, we present an overview of the related work on this topic. This is followed in section 3 by describing the methodology and model specification. This includes a multi-agent transport model, charging stations' placement algorithms, scenarios and evaluation criteria. In section 4, the results of the simulations for the case study are discussed. Finally, in section 5, broad conclusions from the analytical framework and case study results are presented and suggestions for further work are given.

## 2. Prior research

Simulating SAV services and analyzing fleet performance in terms of response times, empty distances, vehicle occupancy rates and more have been done in several previous research efforts. A limited attention, however, has been given to the SAEV fleets. Particularly, a major part of nowadays investigations is focused on the optimization of SAEV infrastructure in which the demand is assumed to be deterministic. In one of the first efforts, Chen et al. (2016) tried to examine the operation of individual ride SAEVs under various vehicle range and charging time scenarios for the case study of Austin, Texas, applying an agent-based simulation built from a former study (Fagnant and Kockelman, 2014). This investigation is based on a spatially aggregated demand model, which is not responsive to the congestion (no traffic assignment or network loading takes place). For the purposes of determining the number of charging stations and the required fleet, passenger access times are integrated into the model and it is assumed that the requests with waiting times exceeding 30 minutes would be eliminated. Once the charging stations and the initial fleet size are determined according to the greedy algorithm, different scenarios with four vehicle ranges and two recharging times are simulated. The simulation results show significant impacts of charging infrastructure and vehicle range on fleet size. It is also suggested that additional vehicle mileages due to accessing charging stations remain less than 5%, with the worst case for the minimum vehicle range and rapid charge scenario. In this study, in each scenario a different number of electric vehicle supply equipment (EVSE) outlets or charging station space is presumed.

Considering the size of fleet, a wide range of vehicle/outlet ratio (the number of vehicles per EVSE outlet) is implicitly assumed (1.6~13.7). As this variation is not applied to the similar scenarios, no conclusion on the impacts of charging space on SAEV service performance could be provided. Based on the mentioned study, Farhan and Chen (2018) tried to evaluate the performance of SAEVs for a ridesharing service. In the extended simulations, a rideshare matching optimization model is proposed to determine optimal routes to pick up and drop off multiple travelers within a given time interval. Two ranges (short and long) and charging speeds as well as four vehicle capacities are assumed for the SAEV fleet. Their results suggest that enabling ridesharing strategy leads to a smaller fleet size and lower number of required charging stations. By switching from individual ride to ridesharing, they realized that the greatest change is occurred when the second passenger is allowed to the vehicle. In ITF (2015) an agent-based model relying on a static representation of the traffic environment is applied to simulate a citywide implementation of SAVs. This study covers scenarios that are more diverse and includes SAEVs. The results suggest that by assuming a fleet of fully electric vehicles equipped with rapid charging batteries (30 minutes) and a range of 175 kilometers, the change on required fleet size remains minimal (+2%). Iacobucci et al. (2019) focused on optimization of SAEV operations upon the transportation network of Tokyo considering charge scheduling and vehicle-to-grid based on the stochastic demand and simplified time-varying traffic stats. The employed optimization includes minimization of wait times and charging costs incorporating dynamic electricity pricing. Their simulation results reveal that while the cost of charging is reduced up to 10%, traveler wait times are not significantly affected by the charging optimization. In the mentioned study, it is however assumed that charging outlets are always available and therefore there are no impacts on service availability and traveler wait times. Kang et al. (2016) developed optimization frameworks for SAEV fleet assignment, charging station placement, and powertrain design. Their proposed service design process is based on a system-level profit-maximization problem according to a long list of operational-, and demand-related decision variables. The demand in their study is however estimated using a marketing approach, which is not responsive to the transportation network and available modes. They conclude that the abovementioned optimizations result in lower traveler wait times. In two latter studies, the impact of wait times on traveler mode choice decision is neglected. Bauer et al. (2018) used an agent-based model and analyzed the cost, energy, and environmental implications of SAEV service operating in Manhattan. An iterative process was employed to optimize the positioning of charging stations by starting with charging stations of one EVSE outlet unit everywhere and eliminating at each iteration the least used chargers. They found the optimal battery size and number of charging stations to minimize costs through sensitivity analysis. They estimated that SAEV costs would be the lowest with a battery range of 50–90 miles, with either 66 chargers of 11 kW and 44 chargers of 22 kW per square mile. They also concluded that currently available EV ranges would be more than sufficient and that reducing battery range from current levels could result in significant cost savings. However, their results being based on a static demand that is built from taxi trips in New York City may not reflect the real usage pattern of a demand responsive transport system.

While aforementioned studies incorporate predefined demands, there are some other investigations that benefit from a dynamic demand, responsive to the network or/and traffic. Loeb et al. (2018) applied a tour-based model coupled with a widely used multi-agent transport simulation platform (MATSim) to anticipate the required charging stations as well as their sizes and positions, assuming a fleet of SAEVs serving travelers across the Austin, Texas region. The main core of this research is similar to Chen et al. (2016), with the difference that a more realistic demand, responsive to traffic, is used. A set of scenarios including various charging times, fleet sizes, vehicle ranges and numbers of charging stations are simulated. Authors conclude that the number of required stations is highly dependent on vehicle range. However, their simulation results suggest that the number of stations is not sensitive to the fleet sizes and charging times. It is also indicated that the faster charging time and longer range (above 175 km) do not essentially improve user wait times. The same authors in a more recent work added gasoline hybrid-electric vehicle to the SAV and SAEV fleet alternatives and compared the performance of proposed service according to the user response time and financial analysis (Loeb and Kockelman, 2019). In their study, an assumption of maximum accepted waiting time is revised and considered to be based on a probability graph in order to make the demand more realistic. Contrary to the previous study, they found that the fleet of long-range (200 miles) SAEV with rapid charging (30 minutes) equipment is the most profitable scenario among the fully electric fleets. Moreover, they conclude that a fleet of gasoline hybrid-electric vehicle is better compared to fully electric vehicles. A summary of the aforementioned studies is presented in Table 1 stating the methodology of demand modeling and traffic simulation as well as the main features.

**Table 1**
Summary of the selected literature on SAEV service simulation with focus on methodology and main features.

| Author(s), year | Demand / Network traffic | Charging station placement | Vehicle/EVSE outlet ratio | Battery capacity-vehicle range | Charging speed - grid connection |
|---|---|---|---|---|---|
| Chen et al., (2016) | Given / static | Greedy algorithm | 1.9/2.4/2.5/13.3[a] | 64/80/160/200 (mi) | 30/240 (min) |
| Farhan and Chen, (2018) | Given / static | Greedy algorithm | NM | 80/190 (mi) | 45/240 (min) |
| ITF, (2015) | Dynamic / static | NM | NM | 175 (km) | 30 (min) |
| Iacobucci et al., (2019) | Given / time-varying | NM | NM | 50 (kWh) | 20/50 (kW) |
| Kang et al., (2016) | Marketing / static | P-median model | 8.0 (optimal sc.) | 44.8 (kWh) | 56 (min) |
| Bauer et al., (2018) | Given / static | Elimination method | 2.8-3.3/6.5/32.5[a] | 10-200 (mi) | 7/11/22/50 (kW) |
| Loeb et al., (2018) | Dynamic / dynamic | Greedy algorithm | NM | 100-325 (km) | 0-240 (min) |
| Loeb and Kockelman, (2019) | Dynamic / dynamic | Greedy algorithm | NM | 60/200 (mi) | 30/240 (min) |

[a] Estimated based on the data provided by each study.

As seen above, most of earlier approaches investigating the operation of SAEV service are based on the predefined or simplified demand and static network traffic except for the Loeb et al. (2018) and Loeb and Kockelman (2019), both of which limit the SAEV mode choice decision with an acceptance waiting time or trip distance rate (level of services is ignored). Our prevoius work demonstrates that by considering dynamic demands, the service usage changes significantly according to the service configuration (Vosooghi et al., 2019b). In the case of SAEV with limited range, the service is relativeley less available. It is therefore necessary to take into account the demand that is dynamicaly responsive to the network and available alternatives. The charging station placement and its impacts on SAEV service performance are also remained missing components in all of the prior studies. Given the cost of providing such infrastructure especially in the high-density areas, the charging station placement resulting in a different operational metrics can certainly affects the profits. Furthermore, even if most of aforementioned investigations incorporate financial analysis, the variation of EVSE outlet ratio, which is another important parameter on the infrastructure cost estimations, is either omitted.

## 3. Model specification and set up

### 3.1. Simulation framework

The present work is based on our previous investigation of non-electric SAV service design and simulation (Vosooghi et al., 2019b), which employed the multi-agent activity-based simulation MATSim (Horni et al., 2016). In this study, in order to simulate SAEVs, the electric vehicle extension proposed by Bischoff et al. (2019) is furthermore used. For setting up the simulation, a synthetic population of the case study area is generated using fitness-based synthesizing with multilevel controls developed previously (Kamel et al., 2018). Activity chains are extracted from a recent transport survey (EMD 2017) and an analysis of population census data (INSEE 2014), and are allocated to each individual of the synthetic population. Once the transportation system of the case study is simulated and calibrated according to the actual modal splits, the SAEV mode and its users' taste variations in terms of mode choice, based on our previous work (Vosooghi et al., 2019a) and a local survey (Al-Maghraoui, 2019), are integrated into the model. In order to allocate the SAEVs more efficiently, a dispatching algorithm developed by Bischoff and Maciejewski (2016) is used. The vehicle dispatch algorithms are slightly adapted taking into account vehicle's state of charge (SoC) when assigning it to a passenger. As such, a vehicle can only be dispatched if its SoC is sufficient to complete the trip and reach a charging station. Vehicles are sent to nearby charging stations with available charging capacity. Should no charger be available, the vehicle is queued at the closest charger until a spot becomes available. This is an extension of a heuristic originally developed in (Bischoff and Maciejewski, 2014).

The passenger waiting time without any limitation on maximum acceptance value is also integrated to the mode choice model incorporated in MATSim in the form of utility scoring. This simulation produces a more accurate estimation of SAEVs service demand compared to the reviewed studies mentioned in Table 1, as the demand is dynamically responsive to the network and traffic. Furthermore, the simulation is performed in the multimodal network in which users can choose other modes if SAEVs are not available in relatively acceptable access times and cost.

Since a full optimization process regarding the charging station positions using agent-based simulation is computationally expensive, we propose to generate charging station locations in a separate model as a first step.

*3.2. Charging station placement*

Here, the first part of the SAEV simulation generates a base set of charging stations. The employed data are based on the non-electric SAV users' demands and pick-up and drop-off locations already estimated in the previous study (Vosooghi et al., 2019b). In fact, the dispersion of SAV user pick-up locations determines areas with a high potential for SAEV service requests. If the charging stations are located in those areas, SAEVs that have already finished charging process could reach to the nearest requests with lower empty distances, which results also in lower passenger wait times. Similarly, it is likely that the SAEVs that need to be recharged reach the nearest charging stations with low empty distances after having dropped-off users. Therefore, by considering the start and end locations of potential trips, travel times and distances between requests, charging stations and SAEVs' decision point locations (for going to charging stations) will be minimized. In order to perform the optimization process, those locations are aggregated to the predefined cells. We use the following two optimization models to compute charging station locations.

The first optimization problem is inspired by Asamer et al. (2016) who tried to find charging station locations for urban electric taxis, and it is based on the maximal covering location problem (MCLP) (Church and ReVelle, 1974). For this purpose, the case study area is meshed to a set of cells $C$. For each cell, a value of $i \in C$, counting the SAV pick-up and drop-off locations within the cell is assigned. Moreover, the cells that have a direct connection to the cell $i \in C$ are denoted as a set of neighbors $N_i \subseteq C \setminus \{i\}$. If the cell $i$ is selected for placing a charging station, a direct coverage weight of $w_0: 1$ is assigned to it. Otherwise, its weight is equal to zero. If at least one of its neighbor cells is selected for the charging station placement, a neighbor coverage weight of $w_1: 0.5$ is set to the cell. Otherwise, the neighbor coverage weight is set to zero. The number of charging stations is limited by $P$. The aim is to maximize the sum of covered pick-up and drop-off locations' counts, whereas a sum of direct and neighbor coverage weights for each cell remain less or equal to one. This means that if a cell is selected for placing a charging station, the neighboring cells may not have a charging station inside. Thus, we avoid placing charging stations near to each other and we keep them enough dispersed. The model can be written as the following mixed integer program:

$$\max \sum_{i \in C} c_i x_i \quad (1)$$

$$\text{subject to} \quad \sum_{i \in C} y_i \leq P \quad (2)$$

$$x_i \leq w_0 y_i + \sum_{j \in N_i} w_1 y_j \quad \forall_i \in C \quad (3)$$

$$x_i \in \{0, 0.5, 1\} \quad \forall_i \in C \quad (4)$$

$$y_i \in \{0,1\} \quad \forall_i \in C \quad (5)$$

where $x_i$ is the variable that represents the sum of direct and neighbor coverage weights. Binary variables decide if a charging station is located in cell $i$ or not.

Since the neighbor coverage weighting is not adequately indicative for our optimization goal especially in terms of distances, a second model based on the distance between charging stations locations and the center of cells is proposed. This model is based on warehouse allocation problems (P-Median), which has already been applied by Kang et al. (2016) to determine optimal charging station locations in Ann Arbor, Michigan case study. A similar partition of cells $C$ as in the previous problem is used. Cells' centroids $E$ are determined as a set of candidate positions of charging stations. The number of charging stations to locate is determined by $P$. Similar to the previous model, for each cell's centroid, a value of $i \in L$, counting the SAV pick-up and drop-off locations within the cell is assigned. The distance between cell centroid $i$ and centralized counting of cell centroid $j$ is defined by $d_{ij}$. The objective here is to minimize the counting-weighted distance of pick-up or drop-off locations and charging stations, expressed as the following mixed integer program:

$$\min \sum_{i \in C} \sum_{j \in E} c_i d_{ij} x_{ij} \quad (1)$$

$$\text{subject to} \quad \sum_{j \in E} x_{ij} = 1 \quad (2)$$

$$\sum_{j \in E} y_j = P \quad (3)$$

$$x_{ij} \leq y_j \quad \forall_i \in C, \forall_j \in E \quad (4)$$

$$y_j \in \{0,1\} \quad \forall_j \in E \quad (5)$$

$$x_{ij} \in \{0,1\} \quad \forall_i \in C, \forall_j \in E \quad (6)$$

where $x_{ij}$ is the binary variable that decides if centroid $i$ is satisfied by charging station located in the cell $j$ and makes sure that each centroid is served by exactly one charging station. Binary variables decide if a charging station is located in the cell $i$ or not.

*3.3. Simulated scenarios*

Different scenarios are simulated to examine the operation of SAEVs in Rouen Normandie metropolitan area (France) under various charging and battery swapping station placements, types of charging outlets (in terms of charging speed), number of charging units per station (vehicle/EVSE outlet ratios) and SAEV battery capacities. These scenarios are grouped by the optimization strategies that have been used for locating charging and battery swapping stations. In order to compare and evaluate the scenarios, a set of performance metrics for SAEV service, infrastructure and users (including vehicle and passenger mileages, service usage and wait times for both traveler and SAEVs that are waiting for charging stations) are defined. In all scenarios, a fleet of 3000 standard 4-seats SAEVs is integrated into the simulation. We obtained this number of vehicles as the best fleet size of non-electric SAV service with ridesharing in our previous study (Vosooghi et al., 2019b). In the simulations, SAEVs are initially allocated to the first requests from four main depots that are located homogeneously across the case study area. Since we seek to explore the impact of charging station placement, these depots are considered as not used for charging or battery swapping stations during simulations. According to the size of the case study area, the maximum number of stations is limited to 12. The SAEVs are sent to charging or battery swapping stations once the battery capacity is below 20% or when the trip distance for the next request (by prediction) shows that with the actual SoC the task could not be undertaken. The vehicle battery capacities are parameterized according to the Renault Zoe specifications (41 kWh and 50 kWh for Zoe R110 and Zoe second-generation accordingly) (Renault Zoe technical sheet, 2019). The autonomous version of this car is, at the time of writing this article, being used for real experimentation in Rouen Normandie metropolitan area. The charging speeds are assumed according to the corresponding available and provided supply equipment (22 kW in the case of normal charger and 43 kW for rapid charger) (Renault Zoe technical sheet, 2019). It is also considered that SAEVs stay at charging stations equipped with normal chargers until the fully charge state is reached. However, rapid chargers charge up to 80% of battery capacity and after that, SAEVs leave the charging process. SAEVs are discharged based on the energy consumption model (Ohde et al., 2016) that has been set up in this study according to the Renault Zoe specifications.

The price of the service is assumed as 0.4 Euro per kilometer (direct distance) for all scenarios. This service prices are slightly more expensive than private car ride costs in France (0.3 Euro per kilometer - *DG Trésor* 2018) and it is almost similar to the values that have been estimated or concluded in other investigations; for instance, Chen et al. (2016) estimated the price for SAEV from $0.42 to $0.49 per person-trip-mile.

## 4. Case study results

*4.1. Base-case scenario*

A base-case scenario simulation run was conducted without considering any vehicle range limitation (non-electric SAV).

Table 3 illustrates the service and network related indicators. As seen here, the fleet of non-electric SAVs forms 5.3% of modal shares. On average, 50% of vehicles are in-use mode during a given day simulation. The latter includes times when vehicles are going to pick up a client. The empty distance travelled with that purpose presents on average 15% of overall vehicle kilometers travelled (VKT). As the SAV users pay the service per each travelled kilometer (without considering detour distances), the total in-vehicle passenger kilometer travelled (PKT) indicates the operator revenues and goes up to 1.97 million kilometers. On-board occupancy rates by number of passengers (PAX occupancy ratio) illustrates the use of SAV capacity and shows that 67% of VKT has been with the only one passenger on-board. The average SAV driven distance is estimated about 546 kilometers. This indicator suggests that future SAEVs with todays' EV range will necessarily require recharging infrastructures.

**Table 2**
Summary of SAV service metrics for the base-case scenario.

|  | SAV service with unlimited range |
|---|---|
| SAV modal share (%) | 5.3 |
| Average waiting time (min) | 20.7 |
| Average in-vehicle time (min) | 46.0 |
| Average detour time (min) | 6.1 |
| Fleet usage ratio (%) | 50 |
| Empty distance ratio (%) | 15 |
| In-vehicle PKT (km) | 1.97 M |
| 1 PAX ratio (%) | 67 |
| 2 PAX ratio (%) | 26 |
| 3 PAX ratio (%) | 6 |
| 4 PAX ratio (%) | 1 |
| Average driven distance (km) | 546 |
| Max. driven distance (km) | 866 |

*4.2. Selection of charging station locations*

In order to locate charging stations, the pick-up and drop-off points identified from the base-case scenario were used as the potential areas of the SAEV service requests. Fig. 1 shows a heat map of those point locations across the case study area. As seen in this figure, SAV users are picked-up or dropped-off in three main areas where agglomerations of population and facilities are located.

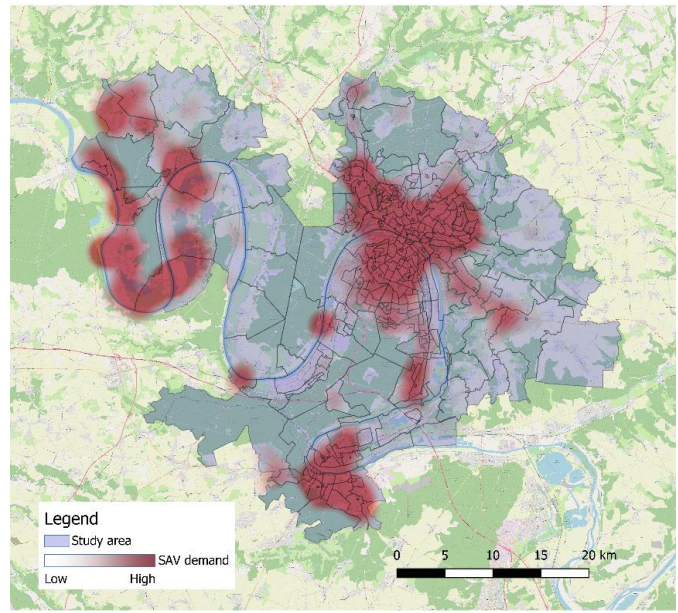

**Fig. 1.** The spatial distribution of SAV demand.

Since we seek to find approximate locations for placing charging stations, in both optimization processes, the SAV pick-up and drop-off points have been spatially aggregated to the uniform cells. Each cell may contain only one charging station. As suggested by Asamer et al. (2016), since a complete tessellation of the study area is required, hexagonal cells are used. The diameter of a hexagon cell is chosen to be one kilometer. For P-Median optimization, the exact location of charging stations is assumed to be in the center of the hexagon. Fig. 2 shows selected hexagons (marked with triangles). As seen here, charging stations are less scattered when MCLP optimization is employed. This occurs because by maximizing coverage location, the distances between demand hubs are somehow neglected. Thus, the charging stations are rather located in the areas where there is a high number of pick-up and drop-off points. The P-Median optimization seeks however the potential locations of charging stations where the access distance from those hubs are minimum. As a result, the selected hexagons are not necessarily where the potential demands are high.

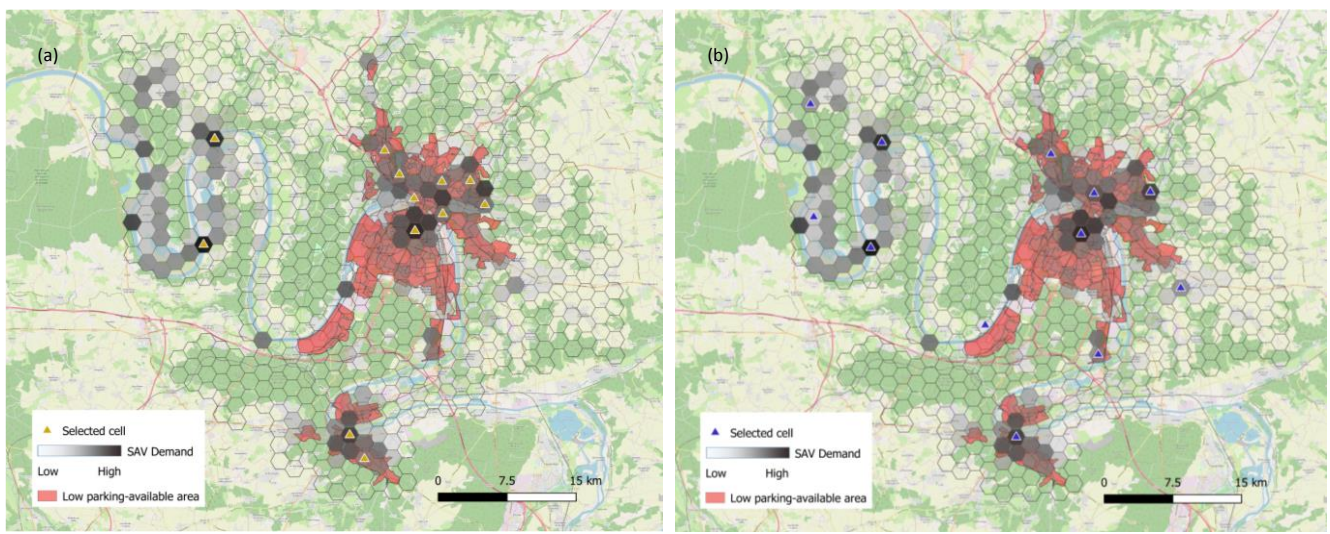

**Fig. 2.** Selected cells for locating charging stations employing different optimization approaches: (a) MCLP; (b) P-Median.

The light red areas superposed in the Fig. 2 show the zones where the parking places are limited. The population and trip attractions in those areas are particularly dense and land values are high. Therefore, in terms of capital expenditures, locating charging stations in areas with low parking availability may lead to excessive cost for the operator. In order to place charging stations outside of these areas we used P-Median optimization with an extra constraint. Fig. 3 shows the selected hexagons. It can clearly be seen that adding this constraint results in a different dispersion of charging stations specifically around the areas with low parking availability. These charging station locations are subsequently evaluated within the dynamic demand multi-agent simulations and compared with the two prior strategies.

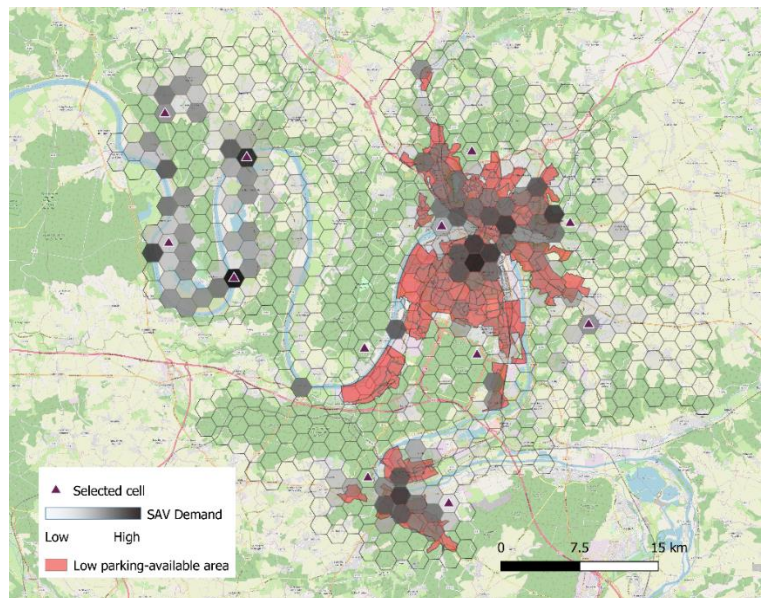

**Fig. 3.** Selected cells for locating charging stations outside the areas with low parking availability employing P-Median optimization.

*4.3. Normal charging infrastructure*

Given the aforementioned charging station placement strategies and two different SAEV battery capacities (41 and 50 kWh), six distinctive scenarios are simulated. Each charging station is assumed to be equipped with 60 outlets of normal charging power (22 kW), which corresponds approximately to the ratio of 4.17 vehicles per EVSE outlet. This ratio is bigger than the one estimated by Chen et al. (2016) for SAEV with normal charge since the vehicle range in that study is shorter. The SAEVs with SoC of less than 20% are dispatched to the nearest charging stations after having dropped-off a client (or clients). The SAEVs stay at charging stations until the battery is fully charged. Fig. 3 shows the result highlighting the SAEV service performance metrics and user related indicators. While simulations are performed for more than 24 hours due to the activity chains that exceed a day, metrics are only calculated for one day. Simulation results show that the SAEV modal shares vary from 3.8% to 4.3% for different charging station placements and battery capacities. The latter remains noticeably lower than non-electric SAV (5.3%, shown in Table 2). Because of the lower service availability due to the charging times along the day, the fleet usages are decreased importantly for all SAEV scenarios compared to the base-case scenario.

The SAEVs perform extra VKT for going to the nearest charging stations, therefore the empty distance ratios are slightly increased compared to non-electric SAVs. Similar to the fleet usage, the in-vehicle PKT decreases considerably and fluctuates significantly for different SAEV scenarios. Considering the latter performance indicator, which presents profits for operator, the strategy of charging station placement outside of areas with low parking availability by minimizing distances between them and potential demand hubs (P-Median with constraint) remains the optimal strategy among all scenarios for both SAEV battery capacities. This occurs because in comparison to the P-Median optimization without location constraint, the charging stations are more dispersed while those with available outlets are accessible within an optimized distance. In fact, given the average trip distance of SAEV users (about 30 km) and size of the case study area, it seems that locating charging stations on the sidelines of potential high demand areas allows the SAEVs to access nearest clients and charging stations (particularly with available outlets) upon an acceptable distance everywhere and when they are. This is also supported by the fact that the maximum coverage location (MCLP) has been shown to be the worse strategy in terms of service performance. In the latter case, the charging stations are rather located in the potential high demand areas close to each other preventing the access of SAEVs that are in "go to charge" mode and that are far from those available charging outlets. This heterogeneity between charging station locations and potential demand hubs affect dramatically the in-vehicle PKT and fleet usage.

For all strategies, the long-range SAEVs perform better in terms of service performance indicators. The impact of battery capacity on service performance is discussed later in this section.

Since charging stations have limited number of outlets, some SAEVs stand in line until charger outlets become available. This could potentially affect the service performance. This metric is actually varying according to different strategies of charging station placement. In order to explore this variation, the total queueing and plugging times of SAEVs are estimated and compared. As seen in

Table **3**, the total queue times are as significant as total plugged times in all scenarios. Only in the P-Median strategy of charging station placement with constraint, the total queue time is slightly less than total plugged time. Given the total plugged time and the number of outlets at each station (60 units), the results show that charger outlets are more efficiently used in the latter strategy. However, the total queue time remains significant and needs to be improved. For this purpose, two main scenarios (the charging stations equipped with rapid chargers and bigger number of outlets per charging station) are simulated. The obtained results will be discussed later in this section.

The user-related indicators do not significantly change amongst all scenarios of charging station placement. The average waiting time varies between 13.2 and 13.9 minutes. Compared to the non-electric SAV, this indicator decreased meaningfully since the service is partially not available during some important times of a given day (especially after peak-hours). Thus, lower SAEV requests by travelers with different trip patterns are served. In all SAEV scenarios, the average detour time fluctuated slightly around 5 minutes. Considering PAX ratios of those scenarios, a correlation between charging station placement and vehicle occupancies is observed. This may occur when there is no strategy of rebalancing. The SAEVs that need to be charged are dispatched to the areas where the spatial trip patterns of travelers are different. In fact, once an SAEV is fully charged, it stays outside of charging station until a request (or some requests) is upcoming. The ride may be shared according to the trip patterns of on-board traveler(s) and next upcoming requests in those areas. As a result, PAX ratios remain almost unchanged for both vehicle ranges for the same charging station locations but vary between scenarios.

**Table 3**
Summary of SAEV service performance and user-related indicators.

| Scenario | MCLP | | P-Median | | P-Median with constraint | |
|---|---|---|---|---|---|---|
| | Medium-Range | Long-Range | Medium-Range | Long-Range | Medium-Range | Long-Range |
| *SAEV* | | | | | | |
| Battery capacity (kWh) | 41 | 50 | 41 | 50 | 41 | 50 |
| Modal share (%) | 3.8 | 4.0 | 4.2 | 4.4 | 4.1 | 4.3 |
| Fleet usage ratio (%) | 31.5 | 34.5 | 36.5 | 38.7 | 35.6 | 41.3 |
| Empty distance ratio (%) | 21.7 | 19.9 | 19.6 | 18.6 | 19.1 | 18.7 |
| In-vehicle PKT (km) | 1.04 M | 1.19 M | 1.13 M | 1.38 M | 1.22 M | 1.44 M |
| Average driven distance (km) | 336 | 365 | 385 | 409 | 373 | 443 |
| Max. driven distance (km) | 660 | 682 | 650 | 698 | 735 | 667 |
| *Charging station* | | | | | | |
| Total plugged time (min) | 381 300 | 399 700 | 433 800 | 451 150 | 443 300 | 496 500 |
| Total queue time (min) | 400 500 | 518 550 | 571 250 | 606 300 | 383 800 | 486 900 |
| *User* | | | | | | |
| Average waiting time (min) | 13.5 | 13.4 | 13.3 | 13.9 | 13.3 | 13.2 |
| Average in-vehicle time (min) | 41.4 | 42.2 | 43.2 | 43.6 | 42.7 | 44.2 |
| Average detour time (min) | 4.7 | 4.9 | 5.0 | 4.8 | 4.9 | 5.3 |
| 1 PAX ratio (%) | 72 | 72 | 67 | 67 | 70 | 70 |
| 2 PAX ratio (%) | 24 | 24 | 28 | 28 | 25 | 25 |
| 3 PAX ratio (%) | 3 | 3 | 4 | 4 | 4 | 4 |
| 4 PAX ratio (%) | <1 | <1 | <1 | <1 | <1 | <1 |

*4.4. Rapid charging infrastructure*

As mentioned before, significant total queue times observed in all SAEV scenarios may strongly affect the performance of services. A solution for reducing those times can be to equip the charging stations with rapid charger (or supercharger) outlets. This EV supply equipment is more expensive for the operators, but could be compensated or even neglected by having greater in-vehicle PKT. In order to explore the impacts of deploying rapid charging infrastructure, the same scenarios are simulated with the charging stations equipped with 43 kW outlets (instead of 22 kW).

Table **3** shows the changes in service performance metrics and charging plugged and queue times. As seen here, fleet usage ratios increased in all scenarios with the maximum values for charging stations

15located according to the maximized coverage. This again indicates that one of the reasons for ineffectiveness of this placement strategy is that the charging stations are not being enough dispersed along the service area. Thus, SAEVs that are in "go to charge" mode but are far from available charging outlets have to wait for charging, and consequently are not efficiently used. Nevertheless, by introducing rapid charging infrastructure, the nearest outlet to each SAEVs becomes available in a faster time.

The empty distance ratios increase in all scenarios except for P-Median strategy without constraint. In the latter case, significant improvements in all service performance metrics are observed. Considering the growth of in-vehicle PKT due to the faster charging time, it is demonstrated that both P-Median strategies perform almost similarly and remain much better than the strategy of maximizing potential-demand coverage. The important improvements on SAEV service metrics of P-Median strategy occurred as introduction of rapid chargers allowed to decrease excessive queue times for the charging outlets located at potentially high demand areas (as seen in

Table **3**). By providing rapid charging infrastructure in those areas, SAEVs become more available at a closer distance to the high demand hubs. Thus, VKTs and in-vehicle PKTs improve. The empty distance ratios of P-Median strategy without constraint are slightly lower compared to the one with constraint, which shows that the charging stations are accessible within relatively lower distances in the former case.

**Table 4**
Summary of SAEV service performance indicators and the changes after deploying rapid charging infrastructures.

| Scenario | MCLP | | P-Median | | P-Median – with constraint | |
|---|---|---|---|---|---|---|
| | Medium-Range | Long-Range | Medium-Range | Long-Range | Medium-Range | Long-Range |
| *SAEV* | | | | | | |
| Fleet usage ratio (%) | 37.5 | 41.2 | 41.4 | 42.7 | 41.6 | 42.3 |
| (relative change) | (+19%) | (+19%) | (+13%) | (+10%) | (+14%) | (+2%) |
| Empty distance ratio (%) | 22.8 | 22.7 | 19.2 | 18.3 | 21.1 | 18.8 |
| (relative change) | (+5%) | (+14%) | (-2%) | (-2%) | (+10%) | (+1%) |
| In-vehicle PKT (km) | 1.24 M | 1.39 M | 1.43 M | 1.56 M | 1.42 M | 1.56 M |
| (relative change) | (+19%) | (+17%) | (+27%) | (+13%) | (+16%) | (+8%) |
| *Charging station* | | | | | | |
| Total plugged time (min) | 212 950 | 226 400 | 229 700 | 240 250 | 245 050 | 242 700 |
| (relative change) | (-44%) | (-43%) | (-47%) | (-47%) | (-45%) | (-51%) |
| Total queue time (min) | 92 100 | 143 650 | 203 950 | 122 650 | 19 700 | 79 150 |
| (relative change) | (-77%) | (-72%) | (-64%) | (-80%) | (-95%) | (-84%) |

*4.5. EVSE outlet units variation*

As seen in Table 4, despite major improvements on service performance indicators of P-Median strategy without constraint, the total queue times remain significant and are the highest among all scenarios. This shows the necessity of having more charging outlets in each station (lower number of vehicles per EVSE outlet). This may actually be as costly as deploying rapid charging infrastructures for service providers, since the land values at those potential high demand areas are accordingly high.

Another solution can be to locate charging stations outside of those areas. In order to explore the impact of vehicle per EVSE outlet variation on SAEV service performance, different capacities of charging stations for both P-Median strategies are simulated. In these scenarios, only normal charging speed (22kW) is considered. Fig. 4 shows the evolutions. As expected, by increasing the number of outlets per charging station, the total charging queue times decrease in all scenarios. This is noticeably because more outlets are available and thus less SAEVs pass the time in queue. The in-vehicle PKTs is however fluctuating around a maximum value. Considering long-range SAEVs, in both strategies of charging station placement, the maximum in-vehicle PKT value is reached when 90 outlets per station (vehicle/EVSE outlet ratio: 2.78) are provided. This indicates that having more charging space and less total charging queue time does not necessarily result in higher service performance, particularly in terms of revenue. This occurs because by providing more outlet units per station, SAEVs are dispatched rather to the nearest charging stations (since in this case the probability of having an available outlet in the nearest charging station is high). Those charging stations are centralized to limited areas and they are situated near to each other particularly in the case of P-Median strategy of charging station placement. As a result, SAEVs are somehow rebalanced and dispatched less dispersedly, and consequently they are accessible with lower level of services. Thus, the service performance indicators and especially in-vehicle PKT decline slightly.

Fig. 5 shows the occupancy rates of charging stations (number of SAEVs charged in each charging station during the given day) for both best-performing and maximum numbers of outlet units in each scenario of charging station placement and vehicle ranges. For medium-range SAEVs of P-Median strategy, since the best number of output units is the maximum one, the occupancy rate is compared with that of lower capacity. As seen in this figure, in accordance with the aforementioned illustration, demands for charging are more spatially dispersed in the case of the best-performing number of charging outlet units for long-range SAEVs (particularly there are less demand for the most occupied charging station). By comparing both strategies of charging station placement, it can be also seen that locating charging stations in areas with low parking availability results in excessive usage of some charging stations (e.g. CS-06 in P-Median without constraint).

Considering medium-range SAEVs, the greatest in-vehicle PKT is reached when 100 outlet units in P-Median strategy are assumed. This indicator is however the highest when 80 outlet units in P-Median strategy with constraint of avoiding locating them in areas with low parking availability are considered. In the former case, compared to the lower charging spaces in the station (90 outlet units per station), for the nearly similar dispersity of demands for charging (see Fig. 5), the lower charging queue time occurs, thus the in-vehicle PKT increases. In the latter case, the limitation of charging station space to 80 units of charging outlets results in a better allocation of SAEVs to the areas where the different trip patterns of users are leading to a higher in-vehicle PKT. Table 5 illustrates this phenomenon. As it is shown, while the average trip distances of SAEV users in other scenarios are nearly similar for both number of outlets per station, the latter decreases importantly when a larger capacity of charging station is assumed for the scenario in question. This occurs because in the P-Median strategy of charging station placement with constraint, charging stations are located on the sidelines of potential high demand areas. Thus, with lower charging outlet units per station, SAEVs that require to be charged are dispatched to a wider set of available charging stations, which are consequently farther from nearest demand hubs. The trips performed by the travelers coming or going to the sideline of potential high demand areas are longer. Thus, while providing service to those travelers without extending total charging queue time during the day, total in-vehicle PKT increases.

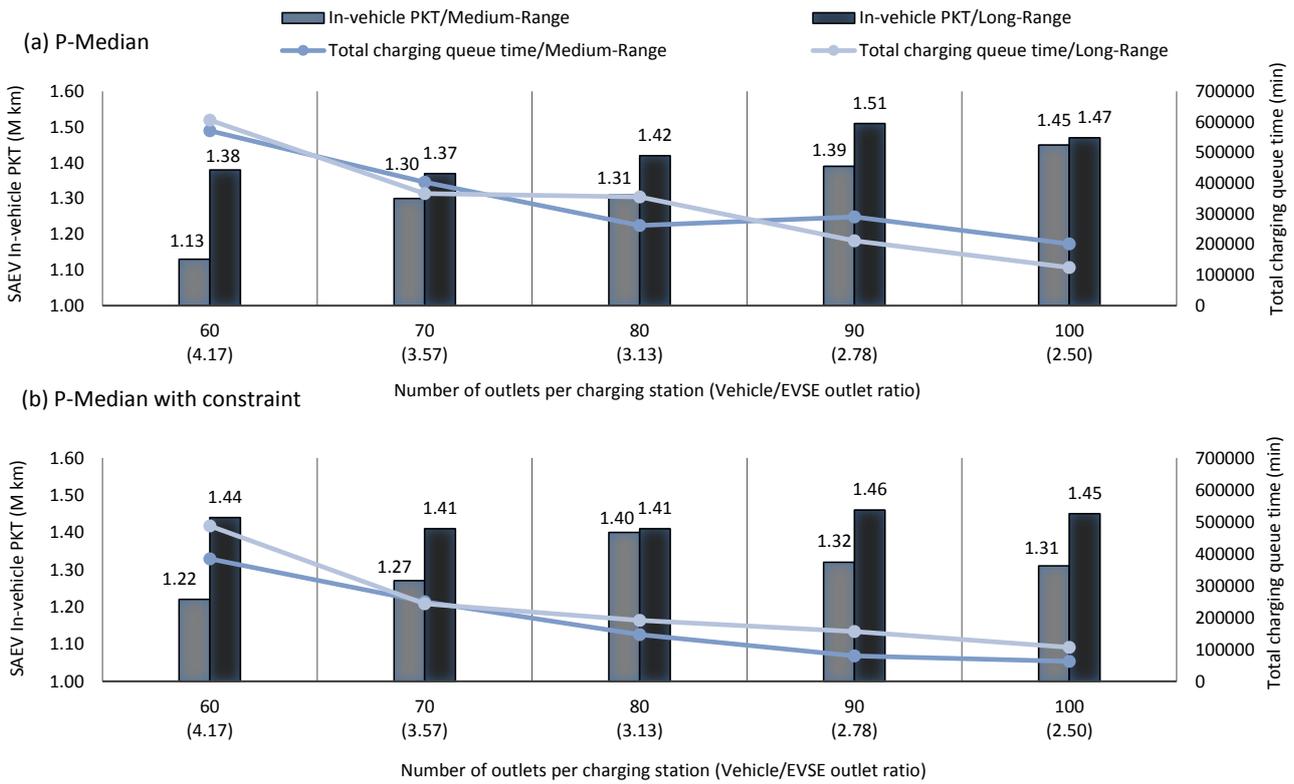

**Fig. 4.** Changes on SAEV service performance indicators according to the different number of outlets per charging station for medium-, and long-range vehicles and two scenarios of charging station placement: (a) P-Median; (b) P-Median with constraint.

**Table 5**
SAEV user average trip distances for different scenario.

| Scenario | P-Median | | | | P-Median – with constraint | | | |
|---|---|---|---|---|---|---|---|---|
| | Medium-Range | | Long-Range | | Medium-Range | | Long-Range | |
| Number of outlets per station | 90 | 100 | 90 | 100 | 80 | 100 | 90 | 100 |
| SAEV user average trip distance (km) | 32.5 | 32.6 | 35.6 | 35.6 | 33.6 | 32.9 | 33.6 | 33.4 |

By comparing the impact of increasing charging outlet units and deploying rapid charging infrastructures, as one can observe in Fig. 4 and Table 4, both policies result in almost similar growth of in-vehicle PKTs particularly in P-Median strategy of charging station placement. This underlines the question, if deploying rapid charging stations is financially more beneficial than providing more charging space. In fact, when the P-Median strategy of charging station placement without any constraint is considered, the latter solution may become costly for the operator. Locating bigger charging stations outside of areas with low parking availability may be however more affordable in terms of capital expenditure. Nevertheless, the final decision has to be made according to the given costs of rapid charging infrastructure and land values as well as the strategies of stakeholders, local authorities and providers of related services (parking, electricity network, etc.).

Another potentially cost-effective solution, particularly for the stations located in the potential high demand areas could be battery swapping (Mak et al., 2013). In this case, the depleted batteries of SAEVs can be exchanged to recharged ones at battery swapping stations (BSS). Since the process is faster than charging, the SAEVs would be more available during the day. Thus, more trips could be served, that result in higher revenue for the operator. However, the additional batteries represent an additional cost for the operator. Therefore, a financial analysis is required to find the best solution. The battery swapping scenario is simulated and the results are discussed later in this section. Due to high uncertainty of future SAEV service and infrastructure costs, in the present study only transport related indicators are evaluated and analyzed.

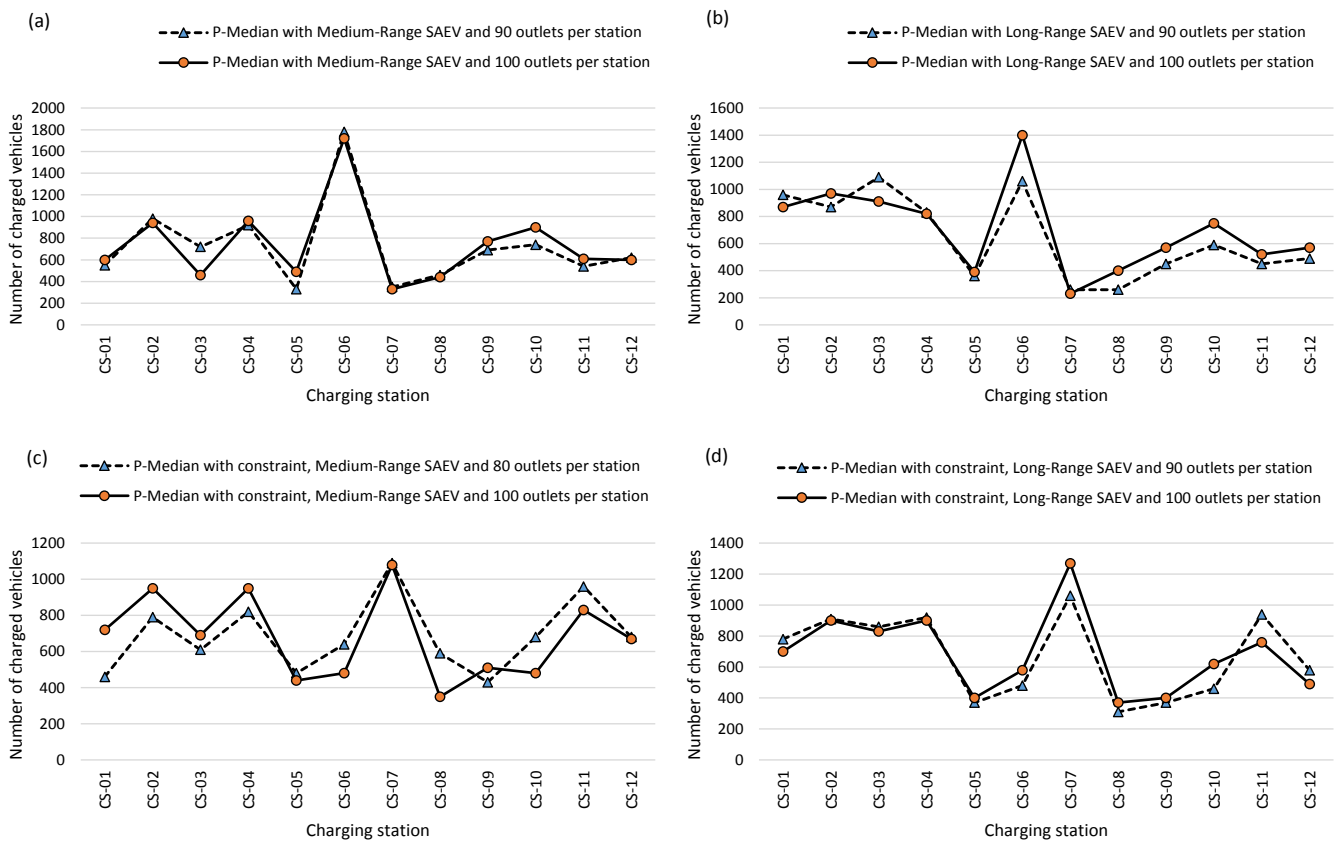

**Fig. 5.** The occupancy rates of charging stations (number of charged SAEVs per charging station during a given day) estimated for the best and maximum numbers of outlets (lower charging outlet units if the best number of units is the maximum one).

*4.6. P-Median strategy with mixed stations*

As stated before, locating spacious charging stations in high-density areas may be very expensive for the service provider. In P-Median strategy of charging station placement, two stations are located in the city center. We reduced the number of outlet units in those stations and increased the capacity of charging stations that are located around the city center (see Fig. 6, the charging station with 10 EVSE outlet units is situated in Rouen Old Town). The overall vehicle/EVSE outlet ratio is supposed to be 2.78, so that we can compare the results with those of previous scenarios. The simulations are done assuming rapid charging in the stations localized in the city center and normal charging in other stations.

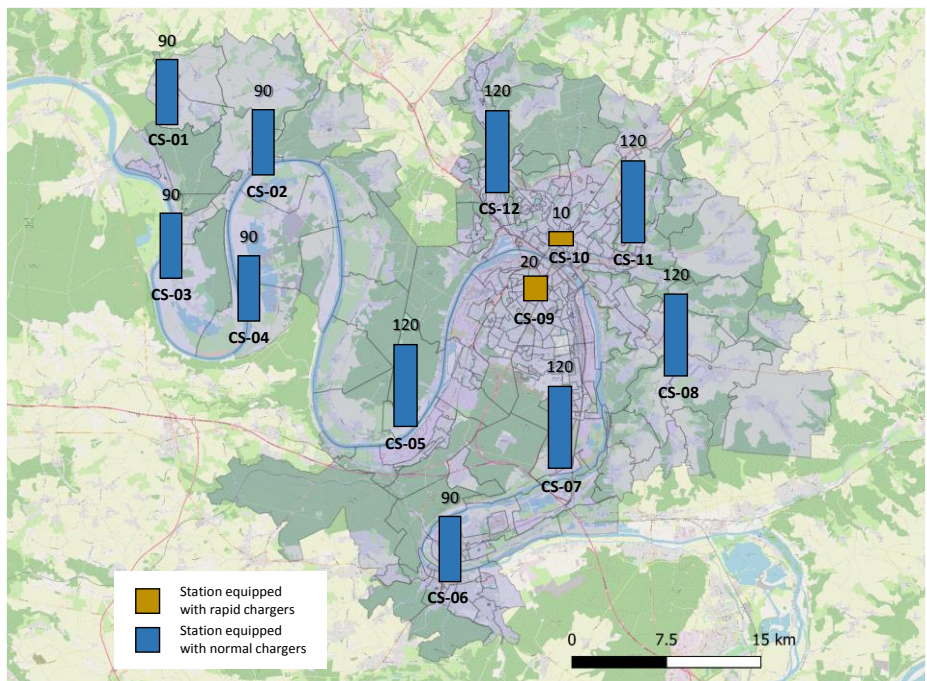

**Fig. 6.** Distribution of charging station outlet units in mixed station scenario.

Table 6 illustrates the results. As seen here, contrary to what was expected, the in-vehicle PKT for both vehicle ranges (battery capacities) decreases. However, the empty distance ratios improve. This occurs since the SAEVs being in the northern regions of the city center area and are in "go to charge" mode have not enough battery to access charging stations located in southern regions. Thus, they wait for the rapid charging stations, in which the number of EVSE outlets are limited, or they try to reach the normal charging stations situated in northern regions, which have already important demands. Consequently, despite the total plugged time that is improved due to the rapid charging, the total queue time increases significantly and those vehicles become less available. In order to avoid high queue time in mixed stations scenario, the limited SoC for going to charge has to be increased or more EVSE outlets in stations located in the city center or northern area should be provided.

**Table 6**
Comparison of SAEV service performance indicators for both P-Median strategies of similar and mixed stations.

| Scenario | P-Median | | P-Median with mixed stations | |
|---|---|---|---|---|
| | Medium-Range | Long-Range | Medium-Range | Long-Range |
| *SAEV* | | | | |
| Empty distance ratio (%) | 19.7 | 19.7 | 19.3 | 18.9 |
| In-vehicle PKT (km) | 1.39 M | 1.51 M | 1.27 M | 1.39 M |
| *Charging station* | | | | |
| Total plugged time (min) | 500 050 | 538 150 | 410 950 | 445 700 |
| Total queue time (min) | 289 750 | 212 400 | 446 750 | 391 700 |

*4.7. SAEV battery capacity (vehicle range)*

The aforementioned simulations incorporate two different battery capacities for each strategy of charging station placement. In all scenarios, the long-range SAEV performs better than medium-range ones in terms of in-vehicle PKTs and empty distance ratios. Fig. 7 shows the SAV hourly in-service rate in base-case scenario illustrating the temporal distribution of potential SAEV demands. The SAEV hourly total plugged times of aforementioned best-performing scenarios (best vehicle/EVSE outlet ratio) using normal charge infrastructures are also shown. As seen in this figure, according to the hourly usage pattern of SAV service, there are two peak hours in a day: morning (8-10) and evening (16-20). Meanwhile, in all scenarios of SAEV, the peak of charging times is after the morning peak hours. In fact, those battery capacities (and accordingly vehicle ranges) are almost enough to meet the demand of morning peak hours. After this time, the majority of medium-range SAEVs face rapidly limited SoC (20%) and are therefore dispatched to the nearest charging stations. In this case, the SAEVs are rather plugged in during the off-peak hours and are ready for service in the evening peak hours. In the case of long-range SAEVs, as mostly SAEVs SoC are enough to meet the demands of midday off-peak hours, they continue to do the service and go to the charging stations later in a day. Thus, important plugged times are rather in the evening peak hours. This actually leads to a lower service performance since a substantial part of SAEVs are not available during evening demand peak hours. As a result, considering the best-performing scenarios of charging outlet units, the difference of in-vehicle PKTs for both SAEV battery capacities are not as significant as expected (see Fig. 4). This actually indicates the importance of battery capacity (vehicle range) and its impact on service performance. Lower SAEV battery capacity may result in missing morning demand. Besides, higher SAEV battery capacity without any change on limited SoC cannot be necessarily beneficial because the operator is obliged to send the SAEVs to the charging stations during midday off-peak hours due to unmatched demand temporal pattern and service availability. However, increasing limited SoC (by more than 20%) in long-range SAEV may result in having more charging station alternatives when a vehicle is in "go to charge" mode as well as avoiding peak charging times in evening peak demand times.

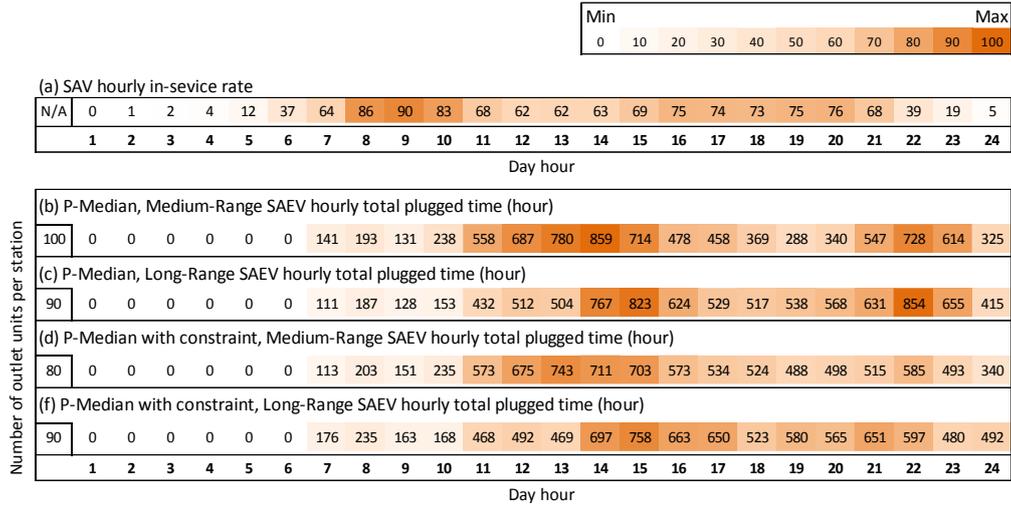

**Fig. 7.** The comparison of SAV hourly in-service rate (base-case scenario) and SAEV hourly total plugged times.

*4.8. SAEV battery swapping*

As shown in Fig. 4, by providing more charging space, the P-Median strategy of charging station placement becomes more efficient in terms of in-vehicle PKT. However, its total charging queue time remains relatively high and still significant. By reducing this time, the SAEV service will be more available and greater in-vehicle PKT can be achieved. A solution may be to provide battery swapping infrastructures. At battery swapping stations (BSS), empty batteries can be replaced by charged ones. This will guarantee that the SAEVs will be supplied with the required energy at the lower unused times (charging and queue). Furthermore, less space would be needed. In order to explore the impacts of battery swapping on SAEV service performance, the same scenarios with BSS located according to the both P-Median strategies were simulated. It is assumed that in each BSS, batteries for 20 SAEVs can be swapped at the same time and the swapping process takes 5 minutes. Table 7 shows the results. As seen here, the in-vehicle PKT increases significantly in all scenarios. The latter remains obviously lower than that of non-electric SAV (1.97 M). Similar to the previous scenarios of SAEV, the empty distance ratios fluctuate around 20% and are bigger than empty distance ratio of non-electric SAV (15%). This is due to the empty drive for going to the BSS during the day, and return back to the initial depots in the end of the day after being fully charged. Even if the battery swapping process takes little time, due to the limited number of BSS spaces and high demand in some areas, the total queue time does not reach zero. The latter is however insignificant for all scenarios (roughly less than one minute per SAEV). Considering in-vehicle PKT and empty distance ratio as the main indicators of service performance, it seems that long-range SAEVs with the P-Median strategy of BSS placement outside of areas with low parking availability is the best-performing scenario. This is the contrary to what we observed previously by increasing charging spaces, where the P-Median strategy with the best number of outlets and without any constraint has a slightly higher in-vehicle PKT. Furthermore, when the medium-range SAEVs are simulated with the BSS infrastructures, the in-vehicle PKT of P-Median scenario performs better. This is actually due to two main reasons. First, since there is no strategy of rebalancing in those scenarios, the location of BSS is somehow affecting the results by distributing SAEVs dissimilarly during the day. As stated before, in the P-Median strategy of charging or battery swapping station placement, those stations are centralized to the potential high demand areas near to each other. Thus, when SAEVs are dispatched to the BSS, they are implicitly rebalanced less dispersedly. As a result, SAEVs become less attractive in terms of access time and consequently the demand decreases (50 410 rides compared to 51 580). This is not however the case of medium-range SAEVs, when the in-vehicle PKT of P-Median strategy is higher. In fact, by deploying BSS infrastructure, SAEVs perform higher distance compared to the previous scenarios (rapid charging and more charging space) especially in morning peak hours. Therefore, they reach rapidly critical SoC and consequently SAEVs are dispatched rather during morning peak hours to the nearest BSS. In the P-Median strategy of BSS placement outside of areas with low parking availability, BSS are accessible with higher distance (this can be actually observed by comparing empty distance ratios). Thus, the service becomes less available during peak hours, which results in lower in-vehicle PKT. This again underlines the importance of battery capacity (vehicle range) and its impacts on SAEV service performance.

Extra battery units required to supply swapping needs are estimated for each scenario. We assume that BSS are equipped with adequate normal chargers (22 kW) and extra batteries are plugged immediately after being detached from SAEV. Once a battery is fully charged, it can be used for the next upcoming request at the same BSS. As seen in Table 7, the number of required extra battery units in all scenarios are less than fleet size (3k SAEVs). This ratio varies between 56% to 78%. This actually occurs since batteries are recharged at a lower time than intervals of two battery swapping for each SAEV. Thus, a battery can be reused for multiple vehicles during a day. Extra battery units needed for the same vehicle ranges in both strategies of BSS placement are almost similar. This indicates the correlation between battery capacities and number of extra batteries. Clearly, the latter varies slightly for each battery capacity according to the total VKT.

**Table 7**
Performance indicators of SAEV service with BSS infrastructure.

| Scenario | P-Median | | P-Median with constraint | |
| --- | --- | --- | --- | --- |
| | Medium-Range | Long-Range | Medium-Range | Long-Range |
| *SAEV* | | | | |
| Fleet usage ratio (%) | 49.9 | 50.9 | 50.1 | 53.0 |
| Empty distance ratio (%) | 20.9 | 19.6 | 21.5 | 19.8 |
| In-vehicle PKT (km) | 1.77 M | 1.82 M | 1.73 M | 1.88 M |
| Total VKT (km) | 1.62 M | 1.64 M | 1.61 M | 1.69 M |
| *BSS* | | | | |
| Total queue time (min) | 2 700 | 3 050 | 1 050 | 1 060 |
| Extra battery (unit) | 2 050 | 2 260 | 1 960 | 2 350 |

## 5. Conclusion

The advent of new shared mobility based on the driverless cars is a widely expected phenomenon in the future. The rising popularity of carsharing in recent years shows important changes that are occurring related to private mobility. Today, people are rather likely to use shared services instead of owning a car. Technological advancements on electric and autonomous vehicles evolving rapidly and the advantages of these vehicles are leading to the emergence of SAEVs. Several car manufactures and transportation network companies have announced their plans for deploying such services in the near future. Providing charging infrastructure is an important prerequisite for this. Present study sought to investigate the impact of charging infrastructure configurations and vehicle's battery capacity on service performance. A multi-agent simulation incorporating dynamic demands responsive to the network, user taste variations and traffic in a multi-modal context was employed. In order to locate charging and battery swapping stations, three strategies of placement were generated in a separated model. These strategies were based on two main optimization approaches: maximizing coverage and minimizing the distance between potential demands and stations. Simulations of non-electric SAVs and SAEVs with two different battery capacities across the Rouen Normandie metropolitan area in France provide initial insights. As suggested by these simulations, future SAEVs with todays' range will necessarily require some recharging infrastructure. Key findings of this research are as following:

First, by providing one normal charger (22 kW) per approximately four SAEVs and limiting vehicle range according to the battery capacities of an autonomous EV used for the experimentation (Renault Zoe, 41 and 50 kWh), the simulations show that the performance indicators are getting dramatically worse in all scenarios compared to non-electric SAV service. Particularly, significant decreases on in-vehicle PKT (27-47% depending on charging placement and vehicle range) that is an indicator of direct revenue for operator and important growths on empty VKT (~34%) are observed.

Second, after replacing normal chargers with rapid chargers (43 kW), important improvements on in-vehicle PKT, especially in the case of medium-range SAEVs are observed. However, the total queue times according to such infrastructure configuration are significant in all scenarios.

Third, by increasing the number of EVSE outlets of normal chargers up to 33-67%, it is seen that SAEVs service reaches the best performance level. Nevertheless, we found that these improvements result in almost similar in-vehicle PKT compared to the case when the rapid charging infrastructure was outspread.

Forth, it is discovered that by providing much lower capacity of battery swapping in each station (20 units) and unlimited normal charge outlets, up to 88-95% of initial in-vehicle PKT (unlimited-range SAV) may be achieved. In this case, extra number of batteries (56-78% more than the number of vehicles) makes SAEV service capable to have vehicles ready to satisfy requests at the minimum time (5 min). Given the service performance indicators of battery swapping simulations, we found that this charging infrastructure is the best alternative among all scenarios. Moreover, by requiring lower charging station capacity and given the feasibility of deploying this infrastructure due to unified vehicles and batteries, the battery swapping station may have a great potential for the providers of future SAEV services.

Fifth, importantly, the choice of charging and battery swapping station placement strategy is found to have a profound effect on service performance indicators. In general, locating charging infrastructure by minimizing distances between potential demands and charging stations leads to much better in-vehicle PKT than when the coverage maximization is employed. It is also observed that the centralization and lower dispersity of charging stations in the low number of charging stations per SAVEs (approximately one unit per four vehicles) may result in the decline of service performance indicators. Further analysis shows that when battery swapping infrastructure are provided the P-Median strategy of BSS placement is the best strategy.

Sixth, increasing charging outlet units in each station result in different impacts depending on the strategy of charging station placement and vehicle range. Two main observations in this case are: a) providing more charging station places does not necessarily lead to better performance indicators, especially when the SAEVs are dispatched to the nearest charging stations without considering the distances from their locations to the upcoming demand hubs, and b) charging infrastructure configuration must take into account the spatial dispersion of charging station usage.

Seventh, the results reveal that the battery capacity of SAEVs has to be set according to the travelled distances of morning peak hours and limited SoC for sending vehicles to charging stations so that maximum charging times occur during midday off-peak hours.

While these investigations and results show significant impacts of charging infrastructure on the SAEV service performance, several other aspects are open to investigation in future work. For example, rather than having the same number of charging spaces or the same charging speed in all stations, future efforts should examine potential combinations of normal and rapid charging as well as different number of outlets in the stations. The authors further plan to integrate a dispatching strategy for the allocation of accessible charging stations to each SAEV within this simulation framework. Understanding financial tradeoff between service benefits (coming from passenger kilometer travelled by SAEV) and charging infrastructure configuration is another important prerequisite for delivering SAEV service, which the authors seek to investigate in the future work.

**Acknowledgements**

This research work has been carried out in the framework of IRT SystemX, Paris-Saclay, France, and therefore granted with public funds within the scope of the French Program "Investissements d'Avenir". The authors would like to thank Groupe Renault for partially financing this work and Métropole Rouen-Normandie for providing the data.